\documentclass[preprint2]{aastex}

\newcommand{\ebv}{E(B$-$V)}
\newcommand{\kms}{km s$^{-1}$}

\newcommand{\mass}{$M_\odot$}

\newcommand{\ha}{H$\alpha$}

\shorttitle{Arp 305}
\shortauthors{Hancock {\it et al.}}

\begin{document}

\title{Candidate Tidal Dwarf Galaxies in Arp 305: Lessons on Dwarf Detachment and Globular Cluster Formation}

\author{Mark Hancock\altaffilmark{1,3}, Beverly J. Smith\altaffilmark{1}, Curtis Struck\altaffilmark{2}, Mark L. Giroux\altaffilmark{1}, \& Sabrina Hurlock\altaffilmark{1}}
\email{mhancock@ucr.edu, smithbj@etsu.edu, curt@iastate.edu, girouxm@etsu.edu, \& zshh7@goldmail.etsu.edu}

\altaffiltext{1}{Department of Physics and Astronomy, East Tennessee State University, Box 70652, Johnson City, TN 37614}
\altaffiltext{2}{Department of Physics and Astronomy, Iowa State University, Ames IA 50011}
\altaffiltext{3}{Now at the Institute of Geophysics and Planetary Physics, University of California, Riverside, Riverside, CA 92521}

\begin{abstract}

To search for Tidal Dwarf Galaxies (TDGs) and to study star formation
in tidal features, we are conducting a large UV imaging survey of
interacting galaxies selected from the Arp\ (1996) Atlas using the
GALEX telescope.  As part of that study, we present a GALEX UV and
SDSS and SARA optical study of the gas-rich interacting galaxy pair
Arp 305 (NGC 4016/7).   The GALEX UV data reveal much extended diffuse
UV emission and star formation outside the disks.  This includes a
luminous star forming region between the two galaxies, and a number of
such regions in tidal tails.  We have identified 45 young star forming
clumps in Arp 305, including several TDG candidates.   By comparing
the UV and optical colors to population synthesis models, we
determined that the clumps  are very young, with several having ages
$\sim6$ Myr.  We do not find many intermediate age clumps in spite of
the fact that the last closest encounter was about 300 Myr ago.  We
have used a smooth particle hydrodynamics code to model the
interaction and determine the fate of the star clusters and candidate
TDGs.

\end{abstract}


\section{INTRODUCTION}

The so-called `Tidal Dwarf Galaxies' (TDG), concentrations of
interstellar gas and stars in the tidal features of interacting
galaxies that may become independent dwarf galaxies, have been the
subject of intense scrutiny (e.g., \citealp{bou06,rec06,duc06}).
There is much uncertainty, at present, about how these structures
compare with other dwarfs, in terms of their interstellar matter,
their stellar populations, their dark matter content, and their star
formation  properties.  Furthermore, the existence of such tidal dwarf
galaxies has been controversial, in part because of questions about
whether they will eventually become independent galaxies.  The
`missing link' that will unambiguously prove the tidal dwarf
hypothesis is the discovery of independent dwarf galaxies that are
detached from other galaxies, but have clear tidal histories.

As part of a search for TDGs and to study star formation in tidal
features, we have used the Galaxy Evolution Explorer (GALEX) telescope
\citep{mar05} to conduct a large UV imaging survey of interacting
galaxies selected from the Arp\ (1996) Atlas \citep{gir09}.   Tidal
features are often very bright in the UV compared to the optical
(\citealp{nef05,han07}).   We have found a number of  previously
unstudied candidate TDGs in our sample \citep{gir09}.

In this paper we investigate the interacting galaxy pair Arp 305 (NGC
4016/7).  We have obtained UV and optical images of Arp 305 from the
GALEX, Sloan Digitized Sky Survey (SDSS), and the Southeastern
Association for Research in Astronomy (SARA) telescopes.  The UV
images reveal extreme amounts of star formation in many different and
well separated environments.

The present paper is organized as follows.  In \S2 we describe the
observations, data reductions, morphology, and photometry, in \S3 we
describe our age and reddening estimates, and discuss the TDG
candidates.  We present a numerical model of the interaction in
\S4, and finally, we summarize in \S5.

\section{OBSERVATIONS}

\subsection{Observations and Reductions}

Arp 305 was imaged with GALEX in both the far-UV (FUV) and near-UV
(NUV) bands covering the wavelength ranges 1350\AA\ to 1750\AA\ and
1750\AA\ to 2800\AA, respectively.  The total integration times in the
FUV and NUV filters were 1584 seconds.  The GALEX images have a 1.2
degree diameter, 5\arcsec\ resolution and a pixel size of 1\farcs5.
The GALEX images were reduced and calibrated through the GALEX
pipeline.

The optical images were obtained as part of the SDSS \citep{aba03}  and
with the SARA telescope\footnote{http://astro.fit.edu/sara/sara.html}.
The SDSS images were observed with the {\it ugriz} optical filters 
with effective wavelengths 3560\AA, 4680\AA, 6180\AA, 7500\AA, and 8870\AA\ 
respectively.  These SDSS images have pixel sizes of 0\farcs40
and the FWHM point-spread function is $\sim$1\farcs2.

Arp 305 was also observed with the SARA 0.9m optical telescope on 2007
February 16, under clear skies.  We used an Axiom/Apogee 2048x2048 CCD
with binning set to $2\times2$ resulting in a pixel size of
$\sim$1\farcs2 pixel$^{-1}$.  A total of four 600 second exposures
were made with a  broadband R filter along with five 600 second images
in a redshifted \ha\ filter centered at 6640\AA\ with a FWHM of 70\AA.
The SARA data was reduced in the standard way using the Image and
Reduction Analysis Facility (IRAF\footnote{IRAF is distributed by the
National Optical Astronomy Observatories, which are operated by the
Association of Universities for Research in Astronomy, Inc., under
cooperative agreement with the National Science Foundation.})
software.  We used the scaled R band image for continuum subtraction.
The white dwarf Feige 34 was also observed for calibration.

\subsection{Morphology and General Properties}

Arp 305 is a very wide pair with the primary galaxy, NGC 4017, to the
South and the companion, NGC 4016, to the North (Figure 1).
Using velocities from the NASA Extragalactic Database (NED),  and
assuming a Hubble constant of 75 km s$^{-1}$ Mpc$^{-1}$, we find Arp
305 is at a distance of $\sim50$ Mpc.  NGC 4017 is designated as SABbc
(NED) and has been classed as  an `H {\scriptsize II} region nucleus'
galaxy by \citet{dah85}.  It appears nearly face-on, with two tidal
tails, one pointing northwest and one to the southeast.  The UV images
show much extended emission to the northwest and southeast far outside
the main disk.   The primary seems to have an ocular waveform, a
bright oval of star formation shaped like an eyelid (see Figure 2).  
This type of
ocular structure has been observed in other galaxies, e.g., Arp 82
(\citealp{kau97,han07})  and IC 2163  and NGC 2207 \citep{elm06}, and
modeling has shown that it is the result  of large-scale gaseous
shocks from a grazing prograde encounter.

The Northern galaxy in Arp 305, NGC 4016, is designated SBdm (NED) and
shows a misshapen bulge with a dusty disk.  In the inner disk, a
curious figure-eight shape is seen (see Figure 2).  There does not
seem to be any evidence to suggest that this figure-eight is an
accretion feature or a wind flow.  It seems too symmetric in  both the
GALEX and SDSS images to result from either.  We suspect that the
figure-eight formation in NGC 4016 is the result of a bar (see the
third frame of Figure 3 in \citealp{rom08}).  More work is needed
before we can fully understand the dynamic environments and formation
of figure-eight features in induced bars.  If the figure-eight is due
to the bar, then its symmetry suggests that the inclination of the
companion, like that of  the primary, is low.  This agrees with the
presence of long tidal arms.

We find much extended UV emission in these galaxies, particularly
outside the primary disk (Figures 1 and 2).   
Recent GALEX observations have
shown a correlation between FUV emission and HI column density within
the tidal tails of galaxy mergers (e.g. \citealp{nef05,thi05, gil05}).
Comparison of the UV images with the HI images of \citet{moo83}
shows that the UV extends out to the third HI contour ($4.1\times10^{20}$)
implying a star formation threshold of N$_{HI}\sim4\times10^{20}$
atoms cm$^{-2}$.  For comparison, \citet{man07} found that the super
star cluster (SSC) density of eight tidal regions in four merging
galaxies was the highest when the HI column density was
N$_{HI}\gtrsim4\times10^{20}$ atoms cm$^{-2}$.  They also showed that
this threshold is a necessary but not sufficient condition for cluster
formation.

The two disks in Arp 305 are separated by $\sim$356\arcsec\ ($\sim$86
kpc).   Assuming an interaction velocity of 300 \kms, the two galaxies
passed close by each other about 300 Myr ago.  This is consistent
with our numerical modeling (see \S4.2).  The star formation in
the clumps has occurred much more recently than this (see \S3.1
below).

\citet{elm93} noted that this galaxy pair showed scattered debris
resembling dwarf galaxies.  The most prominent debris is seen in a partial
residual bridge between the two spiral galaxies.  This feature is
particularly striking in the GALEX images (Figure 1).  For simplicity,
we will adopt the name `bridge TDG' for the tidal dwarf candidate in
the residual bridge.  With SARA, we have detected \ha\ emission from
this feature confirming that it is at the same redshift as the
galaxies (Figure 2).  This structure is clearly detected in HI
\citep{moo83}, further suggesting that it is part of the Arp 305
system.  HI emission can be seen extending from the main NGC 4017 disk
North toward the companion.

NGC 4016 and NGC 4017 have total masses of 1.7 and
$16.2\times10^{10}$\mass\ and HI masses of 2.1 and
$7.1\times10^{9}$\mass, measured out to $\sim$1\farcm5 and
$\sim$3\farcm5 respectively \citep{moo83}.  In an HI survey of Arp
galaxies, \citet{bus87} give a median HI mass for their sample of
$1.8\times10^9$\mass.

Using the calibration in \citet{jes05} we converted the total g band
luminosity of the two galaxy disks to a total Johnson B luminosity.
NGC 4016 and NGC 4017 have L(B) of $\sim4\times10^{9}$ L$_{\odot}$ and
$\sim9\times10^{9}$ L$_{\odot}$ respectively.  For comparison, the
median L(B) in the \citet{bus87} Arp sample is $2.6\times10^{10}$
L$_{\odot}$.  NGC 4016 and NGC 4017 are fainter in B than 97\% and
16\% respectively, of the \citet{bus87} sample of interacting
disk-type galaxies, thus NGC 4016 is somewhat low luminosity.  NGC
4016 and NGC 4017 have mass/luminosity ratios,  M$_{HI}$/L(B), of 0.5
and 0.8 respectively.  The M$_{HI}$/L(B) of NGC 4016/7 are greater
than 77\% and 92\% of  the \citet{bus87} sample respectively.  Thus
Arp 305 is a relatively  gas rich system.

Our SARA data indicate that NGC 4016/7 have \ha\ luminosities of
$6.4\times10^{40}$ and $1.9\times10^{41}$ erg s$^{-1}$ respectively.
These are similar to the median H-alpha luminosities of the 
\citet{bus87} studies $4.6\times10^{40}$ erg s$^{-1}$.

\subsection{Photometry}

We have identified 45 star forming clumps in Arp 305 in the GALEX FUV
images (Figure 2).  The bridge TDG contains four of these
clumps (labeled clumps 12, 13, 15, and 16 in Figure 2).  Two
additional possible detached TDG candidates northeast and southwest of
NGC 4016 are marked in Figure 2 (clumps 1 and 19).

We performed aperture photometry on these clumps with the IRAF task
PHOT.  For the clumps, we used circular
apertures of radii 5\arcsec\ in all bands.  This corresponds to 3.3
and 12.6 pixels in the GALEX and SDSS images respectively.  For
background subtraction of the GALEX and SDSS data we adopted a 5 pixel
wide annulus with radii of 4 and 14 pixels respectively.  For the
entire bridge TDG, NGC 4016 disk, and NGC 4017 disk we used circular
apertures  of radii 37\farcs5, 60\arcsec, and 82\farcs5 respectively
in all bands.  The clumps in the disks and some of the tidal regions
are often  crowded.  Additionally, the backgrounds in the disks change
rapidly.  The sky values for our photometry were determined from the
median of the sky pixels,  which is useful for computing sky values in
regions with rapidly varying sky backgrounds.  We made a second sky
determination using the mode of the same the same sky  annuli which is
suited for measuring stellar objects in crowded fields.  The
difference in the two resulting magnitudes was added as an additional
source of uncertainty to our measurements.

Because there were no suitable stars in the Arp 305 FUV field we
adopted the aperture corrections used for photometry of clumps in Arp
82 of 1.15 for the FUV and 1.45 for the NUV \citep{han07}.  The UV
fluxes were converted to magnitudes on the AB system \citep{oke90}.

For the SDSS optical data, the aperture corrections were determined by
measuring three stars in the field.  The flux aperture corrections for
the {\it ugriz} SDSS data are 1.08, 1.04, 1.07, 1.12, and 1.03 respectively.

Table 1 lists the total magnitudes for the bridge TDG and the two
disks, where the first column is the object, the second through eighth
columns are the FUV, NUV, u, g, r, i, and z magnitudes respectively.
The clump magnitudes are listed in Tables 2, 3, and 4.   In each
table, the first column is the clump number, the second column
describes the clump (e.g. a tidal feature or a disk clump), the third
through ninth columns are the FUV, NUV, u, g, r, i, and z magnitudes
respectively.

\section{RESULTS}

\subsection{Reddening and Ages}

Several of the star forming clumps in Arp 305, including the four in
the bridge TDG, are very blue suggesting very recent star
formation.  To estimate the ages of the star forming clumps we have
generated a  set of model cluster spectral energy distributions (SED)
using the stellar population synthesis code Starburst99 (SB99)
\citep{lei99}.  We used the v5.1 code with the Padova tracks which
include the asymptotic giant branch (AGB) stellar models
\citep{vaz05}.  The SB99 models were generated assuming stellar and
nebular continuum emission, a single burst population, solar
metallicity, a Kroupa IMF \citep{kro02} from 0.1 to 100 \mass, and a
total mass of $10^{6}$\mass.  The model ages range from 1 Myr to 10
Gyr.  To compare with the observations, the models were reddened from
0$-$2.0 mag in 0.02 mag increments assuming the \citet{cal94}
starburst reddening law.   We have included \ha\ line emission in
our model SEDs, as it can have a significant effect on the colors of
young clusters (e.g. \citealp{smi08}).   Finally, the SEDs were
convolved with the GALEX and SDSS bandpasses.  Once the model IMF and
metal  abundance is defined, the location of the clumps on a
color-color diagram depend only on the age and reddening.

Figure 3 plots the FUV-NUV color against the NUV-g color.  The solid
black squares are clumps in the bridge TDG (clumps 12, 13, 15, and 16), the
open blue squares are clumps in NGC 4016, the solid blue triangles
are clumps 1 and 19, and a green x represents clumps in
NGC 4017.  The curves are SB99 models with \ebv\ from 0 to 0.5
mags.  The model ages increase up and
to the right (on this figure) from 1 Myr to 10 Gyr.  
The other color-color figures are similar so are not shown.

To estimate the ages and reddenings of the clumps, we compared the
available measured colors to each of the reddened
SB99 model colors.    We used a $\chi^2$ calculation to determine the
fit of the observed colors to that of the models and hence the ages of
the clumps.  The ages and extinctions associated with the minimum 
$\chi^2$ are taken as the best-fit values \citep{han08}.  To
determine the uncertainties in the predicted ages we find the minimum
and maximum ages within a $\Delta\chi^2$ defined to give a 68\%
confidence level (e.g., \citealp{pre92}).  Additionally, the model
step sizes were added to the age and \ebv\ uncertainties (see
\citealp{han08,smi08} for more details on estimating ages and age
uncertainties).  We note that, because some of
the clump's masses are possibly low, stochastic sampling of
the IMF could affect the age determinations (e.g., \citealp{cer03}).
We do not include this effect in our calculations.

The ages are listed in Table 5 where the first column is the
clump number, the second column is the clump ID, the third and fourth
columns are the R.A. and Dec., the fifth column is the age estimate,
the sixth column is the estimated \ebv, and the seventh column lists
the colors used to determine the ages.  Note that we did not include 
upper limits in our fits.

From Table 5 we find the mean best-fit age of the clumps in the bridge
TDG is $19^{+36}_{-15}$ Myr.  The two other TDG candidates,  clump 1
and clump 19, have best-fit ages of $236^{+23}_{-26}$ and
$7^{+43}_{-2}$  Myr respectively.   Given the precision of our age
estimates, the results in Table 5 suggest that, with the exception of
clump 1, the most recent star formation occurred in
all the clumps at about the same time, $\lesssim10$ Myr ago.

We do not find many intermediate age clumps in spite of the fact that
the last closest encounter was about 300 Myr ago (see \S2.2 and \S4).
One possible explanation is a selection effect.  We selected the
clumps from the FUV image so have chosen the youngest clumps.  Another
possibility is our limited resolution.  By definition the clumps  are
5\arcsec\ or 1.2 kpc in radius (see \S2.3).  It is likely that the
clumps are made up of several unresolved clusters and/or associations.
The light would be dominated by the younger clusters.

It seems likely that our clumps are clusters of clusters that fill a
significant part of our resolution element.  After 300 Myr some of the
clusters may have dissolved.  More importantly, if a cluster of
clusters (a clump) is not bound, the individual clusters may have
simply separated, so that a resolution element centered on any one of
them would not include the others.  Intermediate age clusters are
likely too faint for our observations.  The absence  of intermediate
age clumps in the tidal structures of Arp 305 and some other systems
(e.g, Arp 82 and Arp 285; \citealp{han07,smi08})  indicates that it is
hard to make long-lived TDGs.

We note that the FUV$-$NUV colors of several clumps (mostly in the
southeast tidal region of NGC 4017) do not match our SB99 models
within the $1\sigma$ uncertainties, in that they are too  blue (Figure
3), however, they are consistent within $2\sigma$.  These clumps might
be foreground stars or distant quasars, however,  they do appear
within the extended diffuse UV emission.   Alternatively, our adopted
uncertainties might be too small since we have not included
uncertainties associated with the aperture corrections.  A third
possibility is that the solar metallicity we have assumed may be too
high for our tidal features.  The GALEX FUV$-$NUV color is bluer for
lower metallicity (e.g. \citealp{boi08}).

\subsection{The Bridge TDG}

The bridge TDG looks to be embedded in a massive HI plume stretching
North  from the primary \citep{moo83}.  Four clumps in the bridge TDG
(clumps 12, 13, 15, and 16) are visible in the GALEX image, but only
two (clumps 12 and 13) are prominent in the \ha\ and SDSS images.  
This suggests that the latter two are younger, have more O-stars,
or less extinction.  Our age and extinction estimates of these clumps are not
precise enough to show such a difference (Table 5).

The star formation in the bridge TDG most likely started after it  was
separated from the host disks.  The distance between
the nucleus of the primary and clump 13 in the bridge TDG is 2\farcm5,
$\sim36$ kpc.  The clumps in the bridge  TDG are an average of 19 Myr
old.  To travel 36 kpc in 19 Myr the material would have to travel at
a velocity of roughly 2000 \kms\ relative to the primary.  The HI
kinematics do not support this \citep{moo83}.

Based on our best-fit models and scaling to the SDSS r flux, we
estimate that the stellar masses for clumps 12, 13, 15, and 16 in the
bridge TDG are  $0.3-2.7\times10^{6}$\mass,
$0.3-2.1\times10^{6}$\mass, $0.1-0.9\times10^{6}$\mass, and
$0.1-1.6\times10^{6}$\mass\ respectively, giving a total stellar mass
for the bridge TDG of $\sim1-7\times10^{6}$\mass.  These clump masses
are consistent with that of Galactic globular clusters (e.g.,
\citealp{pry93}) and  Super Star Clusters (SSC)
(e.g. \citealp{hol92,hol96,oco94,sch96,whi93,whi95,wat96}).   For
comparison, the TDG candidates in \citet{hig06} have masses around
2$\times10^7$\mass\ to 3$\times10^8$\mass.

The HI mass of the bridge TDG, as estimated from the contours in
\citet{moo83}, is $\sim6\times10^{7}$\mass.  consistent with the 29
low mass dwarfs studied in \citet{beg08} ($10.18 -
81.14\times10^6$\mass).  This means that the HI mass of the bridge TDG
is about a factor of 9 to 60 greater than the stellar mass.

From the SDSS magnitudes, we estimate a total blue luminosity for the
bridge TDG of $6.1\times10^7$ L$_{\odot}$, in the range of the
irregular and blue compact dwarfs studied by \citet{hun04}.  
The HI mass/luminosity ratio for the bridge TDG is M$_{HI}$/L(B)$\sim1$
M$_{\odot}$/L$_{\odot}$, similar to those of irregular and compact
blue dwarfs (e.g. \citealp{hun04,pis05,tar05,beg08}).  The bridge TDG
has an \ha\ luminosity of of $3.2\times10^{39}$ erg s$^{-1}$, similar
to the \ha\ luminosities of the dwarfs in \citet{hun04} and the tidal
features studied by \citet{smi01}.

The bridge TDG is considerably more luminous than the star forming
regions seen in the HI bridge in the M81/M82 complex (Arp's Loop)
(e.g. \citealp{dem08,boy01,sun05}).  However, they do have similar HI
column densities.  The M81/M82 bridge has n(HI)
$\sim5-30\times10^{20}$ atoms cm$^{-2}$ \citep{dem08}, while the Arp
305 bridge TDG has  n(HI) $\sim4\times10^{20}$ atoms cm$^{-2}$.  For
comparison, the HI bridge between the Magellanic clouds has typical HI
column densities $\sim10^{20}-10^{21}$ atoms cm$^{-2}$
(\citealp{mul03,mul04}).

\subsection{Other Clumps of Interest}

To the southwest of NGC 4016 is a bright clump, number 1, that appears
to be at the tip of a faint tidal arm.   This clump is bright in both
the GALEX UV and all the SDSS optical bands and is within the extended
HI envelope \citep{moo83}.   It is not detected in  our continuum
subtracted \ha\ images and has an \ha\ luminosity $3\sigma$ upper  limit of
$1.4\times10^{39}$ erg s$^{-1}$.  Clump 1 is the oldest clump in our
sample (Table 5).  We can not rule out the possibility that this
object is a foreground star or a background object.

To the northeast is another possible TDG, clump 19.  The northern tail
of NGC 4016 points toward this clump.  Clump 19 is also  bright in
both the GALEX UV and all the SDSS optical bands, and is not detected
in our continuum subtracted \ha\ images with an $3\sigma$ \ha\
upper limit of $2.2\times10^{39}$ erg s$^{-1}$.  Unfortunately, the HI
figure in \citet{moo83} does not cover the region of the sky  that
includes clump 19.  Figure 2 shows clump 19 to be an extended object
with an appearance similar to an inclined disk.  Given this, the lack
of an \ha\ detection, and the large distance from NGC 4016 ($\sim70$
kpc), we can not rule out the possibility that clump 19 is a
background galaxy.

In the northwest tail of the southern galaxy, NGC 4017, there
is another potential TDG, clump 11 (Figure 2).  This clump is seen
clearly in the UV, optical, and \ha\ maps.  It has an \ha\ luminosity of
$2.4\times10^{39}$ erg s$^{-1}$.

Using the same technique described above we find that clump 1 and
clump 19 have stellar masses of $\sim1.9-2.1\times10^{7}$\mass\ and
$\sim0.6-5\times10^{7}$\mass\ respectively.  These masses are
consistent with those of TDGs and more massive than Galactic globular
clusters.  Clump 11 has a mass of $0.45-3.8\times10^{6}$\mass, similar
to globular clusters and SSCs.

Near the bases of both the northeastern and southwestern tidal tails
in NGC 4017 are extremely luminous clumps (clumps 22 and 32).    These
`hinge clumps' likely form when material is pulled out from  deeper in
the original disk.  This material, gas with  higher initial densities,
is more compressed.  Moreover, being pulled out  in a tail likely
reduces the shear levels of the original disk, allowing self-gravity
to more easily form big clouds.  The `hinge clumps' are the two
brightest UV clumps in the primary and are very bright in \ha.  Clumps
22 and 32 have \ha\ luminosities of $1.1\times10^{40}$  erg s$^{-1}$
and $1.2\times10^{40}$ erg s$^{-1}$ respectively and  account for
about 6\% of the total \ha\ flux of NGC 4017.  A luminous
`hinge clump' was also observed  in Arp 82 at the base of the long
extended northern tail \citep{han07}.

\section{NUMERICAL HYDRODYNAMICAL MODEL OF THE ENCOUNTER}
\subsection{Morphological Constraints}

The long tidal tails and bridges in this system, and the nearly
comparable sizes of the visible galaxies, immediately suggest the
action of strong tidal forces.  The ocular morphology of the NGC 4017
disk, and the possibly collisionally induced bar-like morphology of
the NGC 4016 disk reinforce this impression.  All of these
characteristics are indicative of a prograde fly-by encounter  (see
\citealp{str99}).  The fact
that the disturbances are comparable for both galaxies suggests  that
the interaction was prograde for both and that they are of comparable
mass.  The somewhat more disordered appearance of NGC 4016 may suggest
a slightly lower mass, and perhaps a greater disk inclination
relative to the orbital plane.

These are the primary observational clues the system provides to
constrain models.  Secondary constraints include the detailed wave
structure within the disks.  Another secondary constraint is the large
separation of the galaxies, which together with the very small
velocity difference of the two galaxies \citep{moo83} suggests
that they are bound and near their point of greatest separation.

Because the tidal disturbance is so strong, we suspect that the point
of closest approach was very close.  Yet, because of the lack of
ring-like structures we doubt that the disks inter-penetrated to a large
degree.  Thus, with a good idea of the geometry of the collision, of
closest and farthest separations, of galaxy masses and the current
kinematic interaction age, a dynamical model is quite well
constrained.

\subsection{Model Details}

Both the smooth particle hydrodynamical (SPH) simulation code and  the
collision parameters used to produce the model described below are
very similar to those used in our recent study of the Arp 285 system
\citep{smi08}.  More details on the code are also provided in
\citet{str97}.  We will summarize the defining parameters of the
present model, but not repeat all of the details provided in those
earlier works.

The two model galaxies have rigid, dark halo potentials.  These
potentials have a softened power-law form, such that their rotation
curves rise linearly in a core region and decline very slowly at large
radii (as $r^{-0.1}$).  Both halos have the same form  (Eq. (1) of
\citealp{smi08}) and the same softening length of 3 kpc in the
adopted scaling.  The ratio of the companion to primary halo mass is
0.8.

A total of 42,900 particles were used to model the gas disk of the
primary (corresponding to NGC 4017), and 18,090 were used for the
companion (NGC 4016) gas disk.  Small old star disks were also included
in the model, but will not be discussed here.  The model unit of length
was 1.0 kpc, and the time unit was 200 Myr (as in \citealp{smi08}). 
With these units the initial radii of the gas disks were
10.8 kpc for the primary and 6.0 kpc for the companion.  The latter seems
somewhat small, but is still sufficient to yield an enormous tidal
tail after the encounter (see Figure 4).  The halo scale mass of the
primary (see definition in \citealp{smi08}) is 
M$_{h}=1.5\times10^{10}$ M$_{\odot}$.

The primary disk was initialized in the x-y plane.  The companion disk
is  first set up in the x-y plane, then rotated 30\degr\  around a y-axis
through its center, and then 90\degr\  around the z-axis passing through
its center.  The relative orbit of the companion is in the x-y plane,
so from the point of view of the companion disk, the primary
approaches at a moderately steep angle.  The relative orbit of the
companion is in the same sense as the rotation of the primary, so it
sees the encounter as prograde.  Before it is tilted, the companion
disk orbits in the same (counter-clockwise) sense, so it also sees the
encounter as prograde.  (This is the most significant difference from
the Arp 285 model of \citealp{smi08}).

The initial position of the companion relative to the primary center
is (-8.92, -20., 0.0) kpc.  Its initial relative velocity is (240, 75,
0.0) km s$^{-1}$.  The model includes the effects of a Chandrasekhar type
dynamical friction, as described in \citet{str03}.

Simple optically thin radiative cooling is included, as is feedback
heating from star-forming regions, as described in \citet{str97}.  In
the model a particle turns on star formation when it is cool (e.g.,
below about 8000 K) and exceeds a threshold density value. A constant
threshold value is used throughout the model.  However, it is our
experience that the results do not depend greatly on the precise value
of this threshold.  This is because when the threshold is exceeded, the
local self-gravity included in the model dominates, so any reasonable
threshold value will be exceeded, until the heating is initiated.

\subsection{Model Results}

We should note that our philosophy was to produce an
approximate model as an aid to interpreting the observations, so we
have not run an extensive grid of models in order to produce a very
precise representation of the system.   Nonetheless, the model results
do provide helpful input for interpreting this system and projecting
its future.

Figure 4 and Figure 5 show that the model orbit is quite elliptical
(non-circular), which is required by the constraints discussed in
\S4.1.  The four snapshots of Figure 4 show the overall evolution of the
system from a time just after closest approach, in the first panel, to
the onset of merger in the final panel.  The two intermediate times
shown in the second and third panels bracket the present, indicating
that about 300 $-$ 400 Myr has passed since the point of closest
approach.  Because the disk of the primary is in the
plane of the relative orbit, it is strongly affected by the encounter,
and produces a strong bridge and counter-tail, shown by red dots in
Figure 4.  This is in general agreement with the HI observations.  The
model suggests that essentially all of the bridge material originates
in the primary.

Despite its tilt relative to the orbital plane, the companion
(corresponding to NGC 4016), produces a strong counter-tail (shown by
green dots).  It produces only a very short bridge component. These
features are in general, but not complete, agreement with
observation.  The position of clump 1 suggests a longer
southern extension.  On the other hand, the HI data show no long gas
plume to the South.  The HI map does show a feature to the northeast of
NGC 4016, and clump 19 is also found in that area.  These
features support the existence of a strong counter-tail.  However, the
model tail is in the northwest, not the northeast.  If the second and
third panels of the Figure 4 were rotated by about 30\degr\ then the
positions of the galaxies would agree better with the observations,
and the tail would lie more to the North.  Nonetheless, somewhat
different values of the collision parameters or companion potential
parameters would be needed to put the tail in the northeast.

The last three panels of Figure 4 illustrate the star formation (SF)
characteristics of the model.  First of all it is clear that SF is
maintained in the centers of both galaxies for the duration of the
encounter. (This is confirmed by snapshots at other times.)  This is
not surprising since prograde flybys subtract orbital angular momentum
from material in parts of the affected disk, just as they add it to
the sectors that generate bridges and tails.  This leads to an
extended period of central compression as this material moves
inward.

Our data suggest that  clumps with ages greater than 20 Myr are rare
in Arp 305.  The model shows SF in all the tidal features from the
time of their formation to the present.  However, the (model) SF rate
is low in the bridge.  The small number of clusters observed there and
their age distribution are in qualitative accord with that result.
Interestingly, in the model SF occurs in the middle of the bridge and
at the bridge base near the primary, but there is gap in between and
towards the companion, as in the observations. This is seen over a
range of times, but the small numbers of SF particles in the model
preclude firm conclusions.  Material balanced between the two galaxies
in the middle of the bridge may be better able to pull together than
that in the radially sheared parts of the bridge that are falling onto
one of the galaxies.

Also in accord with the observations, SF is relatively strong in the
primary counter-tail (Southern tail of NGC 4017) and weak in that of
the companion (region northeast of NGC 4016).  The latter is tidally
sheared to a much greater length than the former; the South primary
tail stays much closer to its parent disk.  Thus, we expect the gas
surface density to be greater in the Southern tail.

Finally, the model does not account for SF to the northwest of the
primary disk, in the vicinity of clumps 11, 14, and 42 in the
observations.  We speculate that this is material splashed out of the
disks at closest approach.  If the distance of closest approach was
only a little smaller, the model would have more of this `splash'
material.

Nonetheless, the model does a good job in accounting for the overall
SF characteristics of the system.  We again emphasize how similar this
system is to the Arp 285 system.  The modeling suggests that the
primary difference is that the encounter was not as strong (nor
`prograde') for the companion in Arp 285, but the mass transfer from
the primary was greater.

\subsection{Future Projections: The Fate of the Clusters and Dwarfs}

\subsubsection{Mid-term Futures: Building a Globular Cluster System}

Star-forming particles or groups of particles in the model represent
new star clusters or TDGs.  Previous studies (e.g., of M51, 
\citealp{bas05} and the Antennae, \citealp{fal05}) suggest that
many of these will soon disappear as a result of unbinding due to gas
expulsion and other forms of `infant mortality.'  These processes
cannot be resolved in the present models.

Related questions include: what is the long term fate of the clusters
that do survive, and do any of those that might represent tidal dwarfs
detach in some sense?  We have investigated those questions by running
our models beyond the present, up to the time of merger, and followed
the trajectories of specific star-forming particles as shown in
Figure 5.  We note that the tracked
particles remained gas particles in the model, and thus, subject to
the effects of shocks, in contrast to real star clusters.  This is
generally a small problem, since in most cases the trajectories of
these particles are ballistic in the changing potential of the two
galaxies, and hydrodynamic effects like the collision of two gas
streams is rare before merger.

The upper two panels of Figure 5 illustrate the fate of SF particles in the
bridge.  The upper left panel shows the trajectories of particles that
turned on SF as the bridge was forming.  Although they are formed in a
small region, their trajectories bifurcate.  Some fall promptly into
the primary, others are captured by and orbit the companion.  (The
zig-zag trajectories of the latter represent orbits around the
companion.)  

The particle in the upper right panel turned on SF at a
time and place like that of the bridge TDG in the observations, and might
illustrate the fate of the bridge TDG.  If so, that fate is to be promptly
captured by the companion and carried into the merger.  It seems very
unlikely that TDGs formed in the bridge will detach and survive the
merger.

The lower left panel shows clusters formed in the companion
counter-tail.  The story here is similar to that of the bridge in that
there is a bifurcation between particles that remain bound to the
companion and those that `escape' to a region dominated by the
combined potential.  These latter include particles in the outer tail
with significant radial velocities relative to the center of mass at
the time of SF.  A comparison to the last panel of Figure 4 shows that
they are among the particles scattered around the residual tail after
merger.  Although they are bound to the joint potential, these particles are
detached in the sense that they are not contained within the sphere of
influence of either individual galaxy before merger.  These
trajectories suggest that clump 19 may be similarly detached.

We note that our model results in very few particles near the location
of clump 1.  The few that are found there remain tightly bound to the
companion, with no hope of becoming detached.

The last panel of Figure 5 shows the fate of clusters formed in the
primary counter-tail.  Essentially all are retained by the primary,
though the orbit of at least one is strongly influenced by the
companion.  Nonetheless, the orbits of some of these make excursions
out to radii of tens of kpc.  Clumps 36$-$40 and 43$-$45 may follow
similar orbits and may be future members of the globular cluster or
dwarf satellite system of the merger remnant.   It is possible that
this tidal tail, along with the bridge and the inner part of the
companion tail, is the nursery of an incipient globular cluster
system. This scenario is supported by the recent results of
\citet{bou08}.

\subsubsection{Long-term Future of the Far-Flung Tidal Dwarf Candidates}

The TDG candidates born in the outer tail have very low
angular momentum relative to the merger remnant, so we expect that
they have very radial orbits.  Figure 6 shows the results of an attempt
to learn about their long-term future.  This figure shows the gas
particle distribution at the time of the last panel in Figure 4, when
the merging process is well advanced.  The plus signs show the location
of two of the TDG candidates from the lower left panel of
Figure 5.  Their association with the tail is clear.  The two curves show
projections of their trajectories a couple Gyr into the future.  These
trajectories are analytic approximations.  Specifically, particle
positions at late times were output from the numerical model and used
to fit a p-ellipse curve from \citet{str06}.  These curves have been
shown to give quite accurate representations of orbits in fixed
power-law potentials (\citealp{str06,lyn08}).  We assume
that the potential fluctuations in the final stages of merging are
small, so that the global potential is well approximated as the sum of
the two individual halo potentials.  Both potentials have the same
form, which is a power-law outside a small core radius.

The analytic curves show us that the TDG candidates fall into the
merger remnant with other tail material.  For the two particles shown
the radii of closest approach are 5.3 and 7.4 kpc.  Although these are
only estimates, they suggest that the objects will come close enough
to experience substantial tidal disruption.  This is even more likely
since they have negligible dark matter halos to help hold them
together.  The curves shown in Figure 6 do not include the effects of
dynamical friction, which are likely to be considerable given the
close approach.  Overall we expect these dwarf candidates to be
disrupted or reduced to a small core and left on a smaller orbit after
the first close passage.

In their extensive N-body study of tidal dwarf formation \citet{bou06}
found that 75\% of the dwarf candidates fell back into the
galaxies within a few $\times10^8$ yr.  The remaining 25\% had a typical
lifetime of more than 2 Gyr.  Most of the later formed in the outer
parts of tidal tails.  The outer tail objects in the lower left panel
of Figure 5 are the same kind of relatively long-term survivors.  Even
though their existence may consist of a single excursion into the
outer halo before fall and destruction, that process can take several
Gyr.  We suspect very few live much longer.

Globular clusters on the other hand, can survive close passages of
more than several kpc.  Both observations and models suggest that
cluster formation is quite prolific in this type of interaction, and
dozens could be produced in the bridge and tails.

\section{SUMMARY}

We present UV and optical images of Arp 305 (NGC 4016/7) from the
GALEX,  SDSS, and SARA telescopes.  The primary, NGC 4017, seems to
have an ocular waveform, while the Northern galaxy, NGC 4016, has an
interesting figure-eight structure that may be due to a bar.  We have
found active star formation in various tidal features outside the main
galaxy disks.  A prominent  TDG is seen in a partial residual bridge
between the two spiral galaxies, with a total HI mass of
$6\times10^{7}$\mass\ and a  mass/luminosity ratio of M$_{HI}$ / L(B)
$\sim1$ M$_{\odot}$/L$_{\odot}$.

We have identified 45 young star forming clumps from the GALEX FUV 
image of Arp 305.  By comparing the various UV and optical colors
to SB99 models, we determined that the clumps are very young, with
several having ages $\sim6$ Myr (Table 5).

We do not find many intermediate age clumps in spite of the fact that
the last closest encounter was about 300 Myr ago (see \S2.2 and \S4).
The absence of intermediate age clumps in the tidal structures of Arp
305 and some other systems (e.g, Arp 82 and Arp 285; \citealp{han07,smi08})
indicates that it is very hard to make long-lived TDGs.

The stellar masses of the four clumps in the bridge TDG are between
$\sim10^5$ and $\sim3\times10^6$\mass\ and the total stellar mass of the
bridge TDG is $\sim1-7\times10^{6}$\mass.  Our mass estimates suggest
that the bridge TDG might be a system of young globular clusters or
SSCs.  Clump 1 and clump 19 have implied stellar masses of
$\sim1.9-2.1\times10^{7}$\mass\ and $\sim0.6-5\times10^{7}$\mass, 
if they are at the distance of Arp 305.
These masses are consistent with TDGs and are larger than Galactic
globular clusters.

We have used an SPH code to model the interaction and determine the
fate of the candidate TDGs.  The model does a good job in  accounting
for the overall SF characteristics of the system.  Our model suggests
that the fate of the bridge TDG is to be promptly captured by the
companion and carried into the merger.  It seems very unlikely that
TDGs formed in the bridge will detach and survive the merger.  Our
model resulted in very few particles near the location of clump 1.
The few that are found there remain tightly bound to the companion,
with no hope of becoming detached.  Clump 19 is the best candidate for
a soon-to-detach TDG, if it is at the same redshift as Arp 305.

The present system illustrates how tidal structures can be prolific
producers of the precursors of globular clusters.  The eventual merger
of the galaxies may lead to the production of even more, as in the
well-known Antennae system.  However, these are unlikely to orbit very
far out into the halo, while those produced in the tails have a much
better chance of forming a system of halo globulars.

\acknowledgments

We thank the GALEX and SDSS teams for making this research possible.
GALEX is a NASA Small Explorer mission, developed in cooperation with
the Centre National d'Etudes Spatiales of France and the Korean
Ministry of  Science and Technology.  This research was supported by
NASA  LTSA grant NAG5-13079 and GALEX grant GALEXGI04-0000-0026.
This work has made use of the NASA/IPAC Extragalactic Database (NED),
which is operated by the Jet Propulsion Laboratory, California
Institute of Technology, under contract with NASA.

\newpage
\clearpage
\begin{figure*}[ht]
\begin{center}
\scalebox{0.85}{\rotatebox{0}{\includegraphics{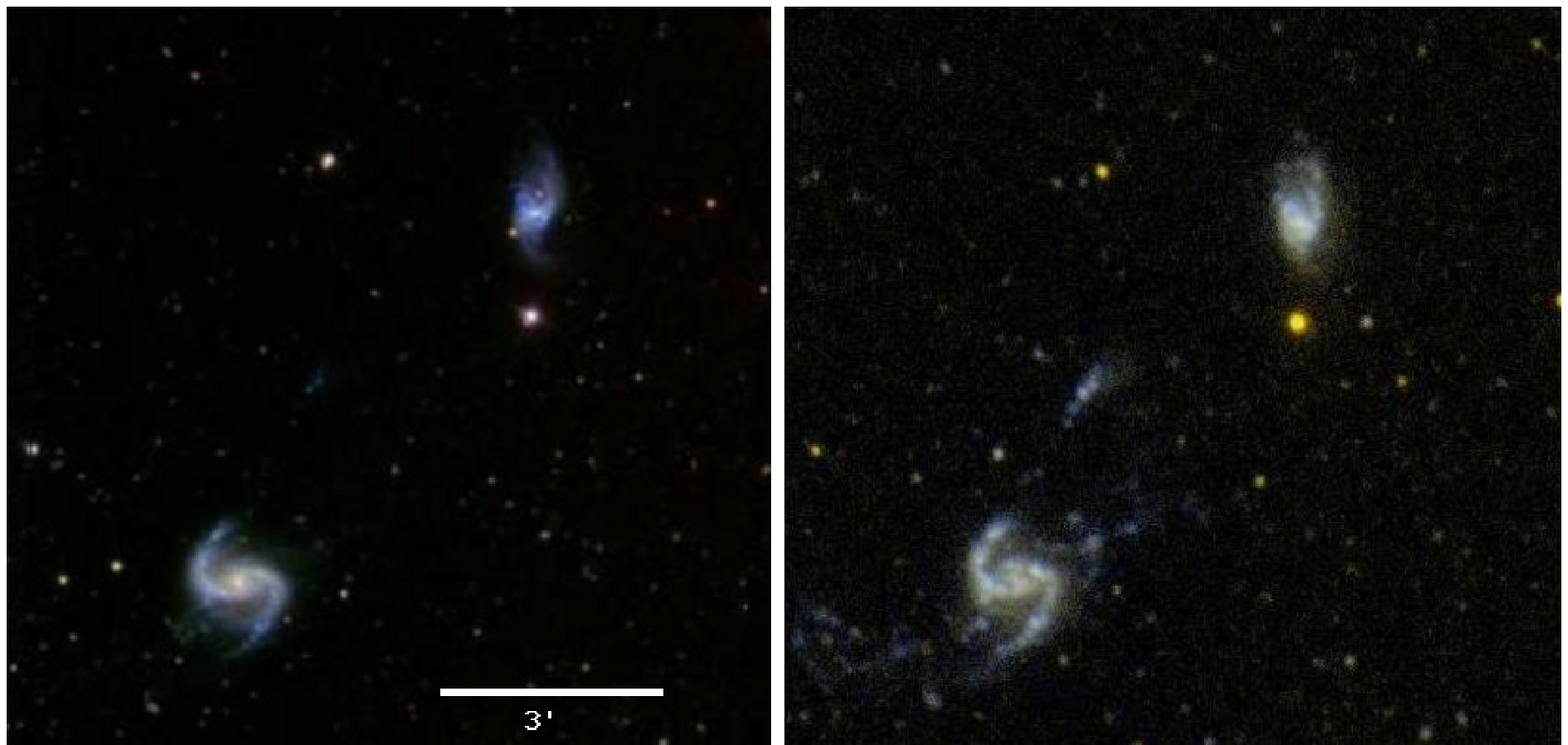}}}
\end{center}
\figcaption[f1.eps]{Left:  Composite {\it ugriz} SDSS image of Arp 305.  North is up and East is to the left.  Right:  Composite FUV (blue) and NUV (yellow) GALEX image of Arp 305.  North is up and East is to the left.  The northern galaxy is NGC 4016, while the southern is NGC 4017.  The bridge TDG is clearly seen between the two galaxies and clump 1 is seen southwest of the NGC 4016 (beside the bright yellow star).  Clump 19 is North of the top of this figure so is not seen.  \label{f1}}
\end{figure*}

\newpage
\clearpage
\begin{figure*}[ht]
\begin{center}
\scalebox{0.55}{\rotatebox{0}{\includegraphics{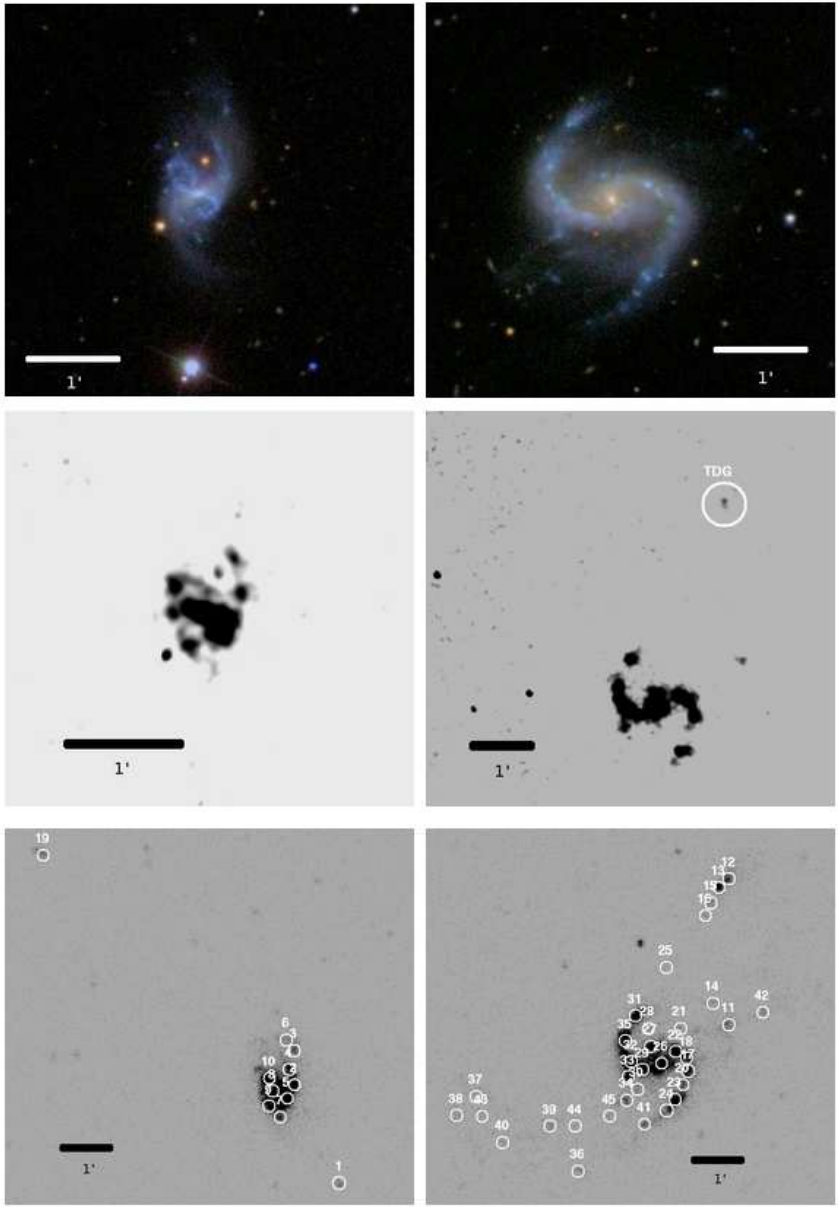}}}
\end{center}
\figcaption[f2.eps]{Top Left:  Composite {\it ugriz} SDSS image of NGC 4016.  North is up and East is to the left.  Notice the striking figure eight shaped feature.  Clump 1 is seen clearly southwest of the galaxy (beside the bright star).  Top Right:  Composite {\it ugriz} SDSS image of NGC 4017.  Middle Left:  Smoothed continuum-subtracted \ha\ image of NGC 4016 obtained with SARA.  The tidal arms and the Northern part of the figure eight features are seen.  Middle Right:  Smoothed continuum-subtracted \ha\ image of NGC 4017 obtained with SARA.  The bridge TDG is clearly detected at the top of this figure.  Bottom Left:  FUV GALEX image of the Northern galaxy NGC 4016 with clumps labeled.  Notice the TDG candidates, one to the South (clump 1) and one to the North (clump 19).  The circles around each clump have radii of 5\arcsec and are the same size as those used for the photometry.  Bottom Right:  FUV GALEX image of the southern galaxy NGC 4017 with clumps labeled.  Notice the TDG candidate to the North (Bridge TDG, clumps 12, 13, 15, and 16).  In each panel North is up and East is to the left. \label{f2}}
\end{figure*}

\newpage
\clearpage
\begin{figure}[ht]
\plotone{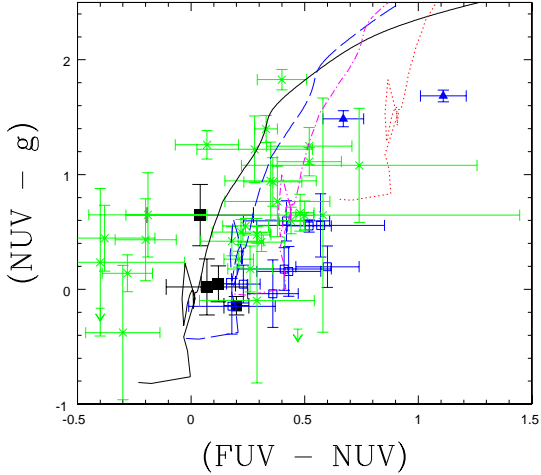}
\figcaption[f3.eps]{FUV-NUV vs NUV-g  color-color plot.  The solid black squares are clumps in the bridge TDG (clumps 12, 13, 15, and 16), the open blue squares are clumps in NGC 4016, the solid blue triangles are clump 1 and clump 19, and a green x represents clumps in NGC 4017.  The curves are SB99 models with \ebv\ of 0 (solid black), 0.12 (long dashed blue), 0.25 (dot dashed magenta), and 0.5 (dotted red).  \label{f3}}
\end{figure}

\begin{figure}[ht]
\plotone{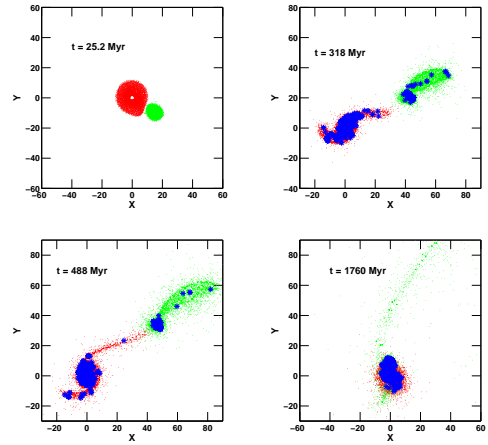}
\figcaption[f4.eps]{Snapshots of the model gas disks from a time near
  closest passage to merger.  The current appearance of Arp 305 best
  matches the views shown in the top right and the lower left panels,
  i.e., $\sim$300 $-$ 400 Myr after closest approach.  Red particles
  originated in the primary galaxy, green in the companion. The top
  left panel shows a time near closest approach. Every fifth particle
  from the simulation is plotted. The axis are in kpc in the adopted
  scaling. In the last three panels blue asterisks mark
  star-forming particles. The companion forms stars actively at most
  times after closest approach. Star formation is common in tidal
  structures at intermediate times. \label{f4}}
\end{figure}

\begin{figure}[ht]
\plotone{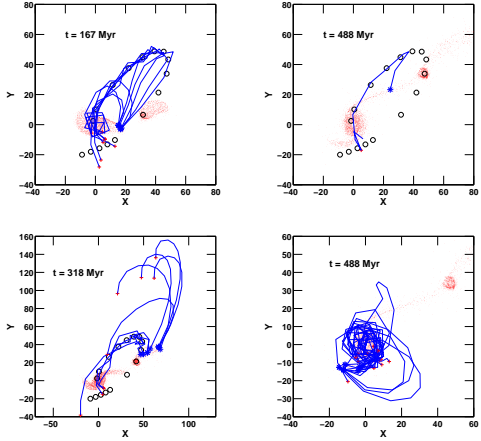}
\figcaption[f5.eps]{Illustrations of the orbital trajectories of
  star-forming particles from the onset of star formation in the
  bridge to the end of the run when the galaxies have merged,
  overplotted on the gas particle distributions at selected times in
  the interaction.  Red dots show the locations of gas particles, at
  the time of SF onset in the selected particles.  Black circles show
  the position of the companion center at selected time-steps from
  closest passage to merger. Blue asterisks show star-forming
  particles in selected tidal structures. The timestep of the panel is
  chosen as the time when these selected particles have just recently
  turned on star formation, so we can follow the subsequent orbit of
  these 'star clusters'.  The axis are in kpc in the adopted scaling.
  Blue curves connecting each asterisk to a red plus sign show the
  particle trajectory from the onset of SF to the end of the run when
  the two galaxies have merged. Note that the particles continue to be
  gas particles, subject to hydrodynamical forces, in contrast to the
  star cluster that would have formed within the
  particle. Nonetheless, it appears that the role of hydrodynamic
  forces is small until such time as the particle collides with a
  galaxy disk. Note also that while the companion galaxy is subject to
  dynamical friction (leading to merger), the individual particles are
  not. Given the small mass of the particles, frictional effects
  should not be large, except for those passing close to the center of
  a galaxy.  Note:  because of the large excursions of the countertail
  particles, the scale of the lower  left panel is larger than that of
  the other panels. \label{f5}}
\end{figure}

\begin{figure}[ht]
\plotone{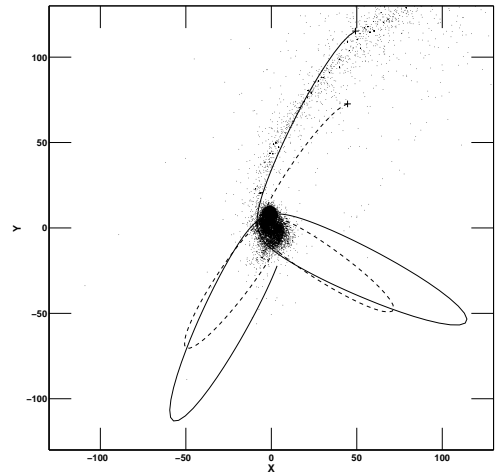}
\figcaption[f6.eps]{Analytic extension of the orbits of two candidate
  tidal dwarf galaxies (solid and dashed curves), calculated as
  described in the text. The dots show the gas distribution at the end
  of the numerical simulation, as in the last panel of Figure 4.
  Crosses mark the start of the extended trajectories, see text for
  more details. The axis are in kpc in the adopted
  scaling. \label{f6}}
\end{figure}

\clearpage
\newpage
\begin{table}[ht]
\caption[fitquality]{Global Magnitudes} 
\vspace{.2in} 
\begin{tabular}{clllllll} 
\hline \hline 
ID & FUV & NUV & u & g & r & i & z \\ 
  & mag & mag & mag & mag & mag & mag & mag \\ 
\hline 
Bridge TDG & 17.90$\pm$0.019 & 17.86$\pm$0.016 & $\geq$19.16 & 17.83$\pm$0.122 & 17.59$\pm$0.175 & 17.73$\pm$0.385 & 17.71$\pm$1.516 \\ 
 NGC 4016 & 15.41$\pm$0.006 & 14.87$\pm$0.003 & 14.97$\pm$0.051 & 14.00$\pm$0.007 & 13.64$\pm$0.009 & 13.43$\pm$0.013 & 13.31$\pm$0.056 \\ 
 NGC 4017 & 14.93$\pm$0.005 & 14.44$\pm$0.003 & 14.07$\pm$0.036 & 13.01$\pm$0.004 & 12.50$\pm$0.005 & 12.25$\pm$0.008 & 12.21$\pm$0.030 \\ 
 
\hline \\ 
\end{tabular} 
\end{table} 
 
\begin{table}[ht]
\caption[fitquality]{Bridge TDG Clump Magnitudes} 
\vspace{.2in} 
\begin{tabular}{cllllllll} 
\hline \hline 
Clump & ID & FUV & NUV & u & g & r & i & z \\ 
  &  & mag & mag & mag & mag & mag & mag & mag \\ 
\hline 
12 & Bridge TDG & 20.0$\pm$0.06 & 19.9$\pm$0.04 & 20.6$\pm$0.36 & 19.9$\pm$0.14 & 19.9$\pm$0.13 & 20.1$\pm$0.22 & $\geq$19.5 \\ 
 13 & Bridge TDG & 19.5$\pm$0.05 & 19.3$\pm$0.02 & 19.5$\pm$0.14 & 19.4$\pm$0.07 & 19.5$\pm$0.10 & 19.6$\pm$0.21 & $\geq$19.5 \\ 
 15 & Bridge TDG & 21.3$\pm$0.10 & 21.2$\pm$0.14 & $\geq$21.3 & 21.2$\pm$0.19 & 21.1$\pm$0.28 & 20.8$\pm$0.42 & $\geq$19.5 \\ 
 16 & Bridge TDG & 21.8$\pm$0.12 & 21.8$\pm$0.20 & $\geq$21.3 & 21.1$\pm$0.17 & 20.9$\pm$0.44 & 20.5$\pm$0.42 & $\geq$19.5 \\ 
 
\hline \\ 
\end{tabular} 
\end{table} 
 
\clearpage
\begin{table}[ht]
\caption[fitquality]{NGC 4016 Clump Magnitudes} 
\vspace{.2in} 
\begin{tabular}{cllllllll} 
\hline \hline 
Clump & ID & FUV & NUV & u & g & r & i & z \\ 
  &  & mag & mag & mag & mag & mag & mag & mag \\ 
\hline 
1 & n4016 tidal & 21.2$\pm$0.09 & 20.1$\pm$0.04 & 19.6$\pm$0.16 & 18.4$\pm$0.02 & 18.6$\pm$0.04 & 18.8$\pm$0.10 & 19.0$\pm$0.48 \\ 
 2 & n4016 disk & 19.2$\pm$0.12 & 18.8$\pm$0.14 & 19.0$\pm$0.36 & 18.2$\pm$0.10 & 18.0$\pm$0.10 & 18.0$\pm$0.14 & 18.5$\pm$0.30 \\ 
 3 & n4016 disk & 20.8$\pm$0.09 & 20.2$\pm$0.09 & 20.2$\pm$0.20 & 20.0$\pm$0.15 & 19.8$\pm$0.11 & 20.2$\pm$0.22 & $\geq$19.5 \\ 
 4 & n4016 disk & 19.7$\pm$0.11 & 19.3$\pm$0.08 & 19.9$\pm$0.26 & 19.1$\pm$0.19 & 19.1$\pm$0.21 & 19.9$\pm$0.24 & $\geq$19.5 \\ 
 5 & n4016 disk & 18.1$\pm$0.13 & 17.7$\pm$0.16 & 17.9$\pm$0.15 & 17.5$\pm$0.09 & 17.2$\pm$0.07 & 17.2$\pm$0.13 & 17.5$\pm$0.42 \\ 
 6 & n4016 disk & 21.1$\pm$0.17 & 20.9$\pm$0.08 & $\geq$21.3 & 21.1$\pm$0.22 & 21.0$\pm$0.28 & $\geq$21.2 & $\geq$19.5 \\ 
 7 & n4016 disk & 19.8$\pm$0.10 & 19.5$\pm$0.04 & 20.1$\pm$0.41 & 19.5$\pm$0.29 & 19.5$\pm$0.33 & 20.5$\pm$0.86 & $\geq$19.5 \\ 
 8 & n4016 disk & 18.3$\pm$0.09 & 17.8$\pm$0.02 & 17.8$\pm$0.10 & 17.3$\pm$0.04 & 16.9$\pm$0.06 & 16.8$\pm$0.14 & 17.0$\pm$0.25 \\ 
 9 & n4016 disk & 19.2$\pm$0.17 & 18.6$\pm$0.22 & 19.1$\pm$0.45 & 18.1$\pm$0.16 & 17.9$\pm$0.19 & 18.0$\pm$0.22 & 19.2$\pm$0.83 \\ 
 10 & n4016 disk & 18.4$\pm$0.05 & 18.1$\pm$0.05 & 18.6$\pm$0.15 & 18.1$\pm$0.15 & 18.1$\pm$0.23 & 18.2$\pm$0.26 & 18.9$\pm$0.34 \\ 
 19 & n4016 tidal & 20.9$\pm$0.08 & 20.3$\pm$0.04 & 19.8$\pm$0.26 & 18.8$\pm$0.05 & 18.3$\pm$0.04 & 18.1$\pm$0.13 & 18.0$\pm$0.22 \\ 
 
\hline \\ 
\multicolumn{9}{l}{\footnotesize{$^a$ Note:  the nucleus is between clumps 5 and 8 (see Figure 5)}}\\ 
\end{tabular} 
\end{table} 
 
\clearpage
\begin{table}[ht]
\caption[fitquality]{NGC 4017 Clump Magnitudes} 
\vspace{.2in} 
\begin{tabular}{cllllllll} 
\hline \hline 
Clump & ID & FUV & NUV & u & g & r & i & z \\ 
  &  & mag & mag & mag & mag & mag & mag & mag \\ 
\hline 
11 & n4017 tidal & 20.7$\pm$0.08 & 20.4$\pm$0.05 & 20.7$\pm$0.33 & 20.0$\pm$0.06 & 20.0$\pm$0.11 & $\geq$21.2 & 19.5$\pm$0.57 \\ 
 14 & n4017 tidal & 21.3$\pm$0.09 & 21.2$\pm$0.10 & $\geq$21.3 & 20.0$\pm$0.06 & 19.4$\pm$0.07 & 19.3$\pm$0.10 & 19.5$\pm$0.53 \\ 
 17 & n4017 disk & 19.6$\pm$0.09 & 19.4$\pm$0.08 & 19.0$\pm$0.08 & 18.9$\pm$0.07 & 18.3$\pm$0.17 & 18.6$\pm$0.19 & 18.2$\pm$0.23 \\ 
 18 & n4017 disk & 19.4$\pm$0.08 & 18.9$\pm$0.11 & 18.4$\pm$0.07 & 17.8$\pm$0.05 & 17.5$\pm$0.07 & 17.3$\pm$0.05 & 17.3$\pm$0.09 \\ 
 20 & n4017 tidal & 20.5$\pm$0.16 & 20.2$\pm$0.19 & 19.9$\pm$0.18 & 19.0$\pm$0.21 & 18.9$\pm$0.15 & 19.0$\pm$0.13 & 18.8$\pm$0.26 \\ 
 21 & n4017 tidal & 21.9$\pm$0.31 & 22.3$\pm$0.19 & $\geq$21.3 & 22.0$\pm$0.61 & $\geq$21.8 & $\geq$21.2 & $\geq$19.5 \\ 
 22 & n4017 hinge & 18.5$\pm$0.03 & 18.0$\pm$0.02 & 17.6$\pm$0.04 & 17.4$\pm$0.12 & 17.1$\pm$0.15 & 17.1$\pm$0.15 & 16.9$\pm$0.09 \\ 
 23 & n4017 tidal & 18.8$\pm$0.03 & 18.6$\pm$0.05 & 18.6$\pm$0.07 & 18.3$\pm$0.04 & 18.2$\pm$0.11 & 18.5$\pm$0.07 & 18.9$\pm$0.42 \\ 
 24 & n4017 tidal & 20.1$\pm$0.08 & 19.8$\pm$0.05 & 20.4$\pm$0.26 & 19.7$\pm$0.19 & 19.7$\pm$0.25 & 19.5$\pm$0.23 & $\geq$19.5 \\ 
 25 & n4017 tidal & 22.3$\pm$0.15 & 22.5$\pm$0.21 & $\geq$21.3 & 21.8$\pm$0.29 & 20.6$\pm$0.20 & 20.0$\pm$0.20 & $\geq$19.5 \\ 
 26 & n4017 nucleus & 18.8$\pm$0.04 & 18.5$\pm$0.02 & 18.0$\pm$0.06 & 17.1$\pm$0.11 & 16.3$\pm$0.19 & 16.0$\pm$0.15 & 15.9$\pm$0.15 \\ 
 27 & n4017 disk & 19.2$\pm$0.17 & 18.8$\pm$0.15 & 19.0$\pm$0.43 & 18.1$\pm$0.26 & 17.6$\pm$0.27 & 17.5$\pm$0.28 & 18.1$\pm$0.92 \\ 
 28 & n4017 tidal & 23.0$\pm$0.57 & 22.5$\pm$0.64 & $\geq$21.3 & 21.8$\pm$0.79 & 21.8$\pm$1.15 & $\geq$21.2 & $\geq$19.5 \\ 
 29 & n4017 disk & 18.9$\pm$0.12 & 18.6$\pm$0.02 & 18.0$\pm$0.10 & 17.6$\pm$0.21 & 17.6$\pm$0.38 & 17.8$\pm$0.50 & 17.3$\pm$0.15 \\ 
 30 & n4017 disk & 22.3$\pm$0.36 & 21.5$\pm$0.36 & 21.2$\pm$0.50 & 20.5$\pm$0.33 & 20.2$\pm$0.37 & 20.2$\pm$0.32 & $\geq$19.5 \\ 
 31 & n4017 tidal & 18.8$\pm$0.04 & 18.5$\pm$0.04 & 18.4$\pm$0.07 & 18.0$\pm$0.05 & 17.9$\pm$0.05 & 18.0$\pm$0.07 & 18.1$\pm$0.14 \\ 
 32 & n4017 hinge & 18.3$\pm$0.08 & 18.1$\pm$0.05 & 18.0$\pm$0.08 & 17.5$\pm$0.03 & 17.2$\pm$0.07 & 17.2$\pm$0.15 & 17.3$\pm$0.12 \\ 
 33 & n4017 disk & 19.2$\pm$0.04 & 18.7$\pm$0.03 & 18.8$\pm$0.20 & 18.1$\pm$0.10 & 17.9$\pm$0.10 & 17.9$\pm$0.10 & 19.3$\pm$0.72 \\ 
 34 & n4017 tidal & 20.5$\pm$0.08 & 20.4$\pm$0.11 & 20.7$\pm$0.43 & 19.9$\pm$0.06 & 19.7$\pm$0.08 & 19.6$\pm$0.15 & 19.5$\pm$0.48 \\ 
 35 & n4017 tidal & 18.9$\pm$0.04 & 18.4$\pm$0.04 & 18.4$\pm$0.14 & 17.7$\pm$0.15 & 17.5$\pm$0.06 & 17.6$\pm$0.17 & 17.6$\pm$0.12 \\ 
 36 & n4017 tidal & 21.3$\pm$0.09 & 20.9$\pm$0.06 & 20.2$\pm$0.19 & 19.0$\pm$0.06 & 18.5$\pm$0.05 & 18.3$\pm$0.05 & 18.2$\pm$0.14 \\ 
 37 & n4017 tidal & 22.1$\pm$0.17 & 21.9$\pm$0.17 & $\geq$21.3 & 21.9$\pm$0.69 & $\geq$21.8 & $\geq$21.2 & $\geq$19.5 \\ 
 38 & n4017 tidal & 21.6$\pm$0.11 & 21.9$\pm$0.12 & $\geq$21.3 & 22.3$\pm$0.57 & 21.5$\pm$0.77 & $\geq$21.2 & 19.3$\pm$0.42 \\ 
 39 & n4017 tidal & 21.4$\pm$0.1 & 21.7$\pm$0.19 & $\geq$21.3 & 21.3$\pm$0.20 & 21.2$\pm$0.30 & 21.0$\pm$0.69 & $\geq$19.5 \\ 
 40 & n4017 tidal & 21.9$\pm$0.12 & 22.3$\pm$0.15 & $\geq$21.3 & $\geq$22.5 & $\geq$21.8 & $\geq$21.2 & $\geq$19.5 \\ 
 41 & n4017 tidal & 20.7$\pm$0.08 & 21.$\pm$0.07 & 20.8$\pm$0.58 & 20.8$\pm$0.14 & 20.8$\pm$0.22 & 21.0$\pm$0.58 & $\geq$19.5 \\ 
 42 & n4017 tidal & 21.2$\pm$0.09 & 21.4$\pm$0.1 & 21.1$\pm$0.49 & 20.9$\pm$0.34 & $\geq$21.8 & 21.2$\pm$999. & $\geq$19.5 \\ 
 43 & n4017 tidal & 22.0$\pm$0.13 & 21.5$\pm$0.13 & $\geq$21.3 & 20.3$\pm$0.08 & 20.1$\pm$0.11 & 18.4$\pm$2.79 & $\geq$19.5 \\ 
 44 & n4017 tidal & 22.4$\pm$0.15 & 22.0$\pm$0.13 & $\geq$21.3 & 21.1$\pm$0.31 & 20.1$\pm$0.11 & 21.2$\pm$999. & 19.5$\pm$0.72 \\ 
 45 & n4017 tidal & 22.6$\pm$0.32 & 22.1$\pm$0.19 & $\geq$21.3 & $\geq$22.5 & 21.2$\pm$0.47 & 19.6$\pm$0.33 & $\geq$19.5 \\ 
 
\hline \\ 
\end{tabular} 
\end{table} 
 
\clearpage
\begin{table}[ht]
\caption[fitquality]{Age and E(B$-$V) Estimates} 
\vspace{.2in} 
\begin{tabular}{clllllc} 
\hline \hline 
Clump & ID & R.A. & DEC & Ages & E(B$-$V) & Colors Used$^a$ \\ 
  &  & (deg) & (deg) & (Myr) & (mag) & \\ 
\hline 
1 & n4016 tidal & 179.60528 & 27.508193 & 236$^{+23}_{-26}$ & 0.00$^{+0.04}_{-0.00}$ & FUV$-$NUV, NUV$-$g, u$-$g, g$-$r, r$-$i, i$-$z \\ 
 2 & n4016 disk & 179.61702 & 27.531111 & 6$^{+85}_{-1}$ & 0.22$^{+0.12}_{-0.22}$ & FUV$-$NUV, NUV$-$g, u$-$g, g$-$r, r$-$i, i$-$z \\ 
 3 & n4016 disk & 179.61702 & 27.539028 & 4$^{+3}_{-1}$ & 0.26$^{+0.12}_{-0.18}$ & FUV$-$NUV, NUV$-$g, u$-$g, g$-$r, r$-$i \\ 
 4 & n4016 disk & 179.61843 & 27.534653 & 5$^{+2}_{-2}$ & 0.22$^{+0.14}_{-0.16}$ & FUV$-$NUV, NUV$-$g, u$-$g, g$-$r, r$-$i \\ 
 5 & n4016 disk & 179.6189 & 27.527778 & 8$^{+24}_{-3}$ & 0.10$^{+0.16}_{-0.10}$ & FUV$-$NUV, NUV$-$g, u$-$g, g$-$r, r$-$i, i$-$z \\ 
 6 & n4016 disk & 179.61914 & 27.541528 & 4$^{+27}_{-1}$ & 0.12$^{+0.12}_{-0.12}$ & FUV$-$NUV, NUV$-$g, g$-$r \\ 
 7 & n4016 disk & 179.62071 & 27.523611 & 4$^{+29}_{-2}$ & 0.20$^{+0.14}_{-0.18}$ & FUV$-$NUV, NUV$-$g, u$-$g, g$-$r, r$-$i \\ 
 8 & n4016 disk & 179.62266 & 27.529653 & 11$^{+16}_{-6}$ & 0.22$^{+0.06}_{-0.10}$ & FUV$-$NUV, NUV$-$g, u$-$g, g$-$r, r$-$i, i$-$z \\ 
 9 & n4016 disk & 179.6236 & 27.526111 & 6$^{+134}_{-3}$ & 0.24$^{+0.22}_{-0.24}$ & FUV$-$NUV, NUV$-$g, u$-$g, g$-$r, r$-$i, i$-$z \\ 
 10 & n4016 disk & 179.6236 & 27.532569 & 6$^{+30}_{-3}$ & 0.10$^{+0.14}_{-0.10}$ & FUV$-$NUV, NUV$-$g, u$-$g, g$-$r, r$-$i, i$-$z \\ 
 11 & n4017 tidal & 179.67126 & 27.460685 & 6$^{+39}_{-1}$ & 0.16$^{+0.06}_{-0.16}$ & FUV$-$NUV, NUV$-$g, u$-$g, g$-$r \\ 
 12 & Bridge TDG & 179.67128 & 27.494644 & 6$^{+32}_{-2}$ & 0.06$^{+0.10}_{-0.06}$ & FUV$-$NUV, NUV$-$g, u$-$g, g$-$r, r$-$i \\ 
 13 & Bridge TDG & 179.67386 & 27.492559 & 15$^{+6}_{-11}$ & 0.04$^{+0.10}_{-0.04}$ & FUV$-$NUV, NUV$-$g, u$-$g, g$-$r, r$-$i \\ 
 14 & n4017 tidal & 179.67549 & 27.465684 & 8$^{+35}_{-2}$ & 0.34$^{+0.08}_{-0.14}$ & FUV$-$NUV, NUV$-$g, g$-$r, r$-$i, i$-$z \\ 
 15 & Bridge TDG & 179.67597 & 27.489017 & 15$^{+27}_{-11}$ & 0.06$^{+0.16}_{-0.06}$ & FUV$-$NUV, NUV$-$g, g$-$r, r$-$i \\ 
 16 & Bridge TDG & 179.67738 & 27.4861 & 39$^{+77}_{-34}$ & 0.04$^{+0.30}_{-0.04}$ & FUV$-$NUV, NUV$-$g, g$-$r, r$-$i \\ 
 17 & n4017 disk & 179.68182 & 27.449848 & 5$^{+7}_{-2}$ & 0.26$^{+0.10}_{-0.12}$ & FUV$-$NUV, NUV$-$g, u$-$g, g$-$r, r$-$i, i$-$z \\ 
 18 & n4017 disk & 179.68229 & 27.452973 & 6$^{+77}_{-1}$ & 0.38$^{+0.08}_{-0.30}$ & FUV$-$NUV, NUV$-$g, u$-$g, g$-$r, r$-$i, i$-$z \\ 
 19 & n4016 tidal & 179.68283 & 27.584431 & 7$^{+43}_{-2}$ & 0.40$^{+0.10}_{-0.14}$ & FUV$-$NUV, NUV$-$g, u$-$g, g$-$r, r$-$i, i$-$z \\ 
 20 & n4017 tidal & 179.68323 & 27.446722 & 140$^{+124}_{-135}$ & 0.00$^{+0.44}_{-0.00}$ & FUV$-$NUV, NUV$-$g, u$-$g, g$-$r, r$-$i, i$-$z \\ 
 21 & n4017 tidal & 179.68394 & 27.459847 & 7$^{+62}_{-6}$ & 0.00$^{+0.30}_{-0.00}$ & FUV$-$NUV, NUV$-$g \\ 
 22 & n4017 hinge & 179.68511 & 27.45443 & 6$^{+19}_{-2}$ & 0.26$^{+0.08}_{-0.06}$ & FUV$-$NUV, NUV$-$g, u$-$g, g$-$r, r$-$i, i$-$z \\ 
 23 & n4017 tidal & 179.68534 & 27.443388 & 6$^{+1}_{-1}$ & 0.12$^{+0.04}_{-0.06}$ & FUV$-$NUV, NUV$-$g, u$-$g, g$-$r, r$-$i, i$-$z \\ 
 24 & n4017 tidal & 179.68745 & 27.440679 & 12$^{+38}_{-8}$ & 0.12$^{+0.14}_{-0.12}$ & FUV$-$NUV, NUV$-$g, u$-$g, g$-$r, r$-$i \\ 
 25 & n4017 tidal & 179.6877 & 27.474012 & 9$^{+24}_{-2}$ & 0.22$^{+0.20}_{-0.22}$ & FUV$-$NUV, NUV$-$g, g$-$r, r$-$i \\ 
 26 & n4017 nucleus & 179.68886 & 27.45172 & 130$^{+84}_{-87}$ & 0.04$^{+0.18}_{-0.04}$ & FUV$-$NUV, NUV$-$g, u$-$g, g$-$r, r$-$i, i$-$z \\ 
 27 & n4017 disk & 179.69168 & 27.455677 & 8$^{+178}_{-5}$ & 0.26$^{+0.24}_{-0.26}$ & FUV$-$NUV, NUV$-$g, u$-$g, g$-$r, r$-$i, i$-$z \\ 
 28 & n4017 tidal & 179.69239 & 27.459843 & 6$^{+382}_{-5}$ & 0.24$^{+0.60}_{-0.24}$ & FUV$-$NUV, NUV$-$g, g$-$r \\ 
 29 & n4017 disk & 179.69379 & 27.450259 & 34$^{+119}_{-30}$ & 0.16$^{+0.24}_{-0.16}$ & FUV$-$NUV, NUV$-$g, u$-$g, g$-$r, r$-$i, i$-$z \\ 
 30 & n4017 disk & 179.6952 & 27.445675 & 6$^{+321}_{-3}$ & 0.38$^{+0.36}_{-0.38}$ & FUV$-$NUV, NUV$-$g, u$-$g, g$-$r, r$-$i \\ 
 31 & n4017 tidal & 179.69568 & 27.462758 & 6$^{+1}_{-1}$ & 0.16$^{+0.06}_{-0.04}$ & FUV$-$NUV, NUV$-$g, u$-$g, g$-$r, r$-$i, i$-$z \\ 
 32 & n4017 hinge & 179.69685 & 27.452132 & 8$^{+5}_{-3}$ & 0.18$^{+0.06}_{-0.06}$ & FUV$-$NUV, NUV$-$g, u$-$g, g$-$r, r$-$i, i$-$z \\ 
 33 & n4017 disk & 179.69755 & 27.44859 & 6$^{+27}_{-1}$ & 0.24$^{+0.06}_{-0.10}$ & FUV$-$NUV, NUV$-$g, u$-$g, g$-$r, r$-$i, i$-$z \\ 
 34 & n4017 tidal & 179.69801 & 27.443173 & 12$^{+31}_{-7}$ & 0.14$^{+0.10}_{-0.14}$ & FUV$-$NUV, NUV$-$g, u$-$g, g$-$r, r$-$i, i$-$z \\ 
 35 & n4017 tidal & 179.69826 & 27.456923 & 6$^{+25}_{-2}$ & 0.26$^{+0.10}_{-0.08}$ & FUV$-$NUV, NUV$-$g, u$-$g, g$-$r, r$-$i, i$-$z \\ 
 36 & n4017 tidal & 179.71068 & 27.426708 & 70$^{+76}_{-64}$ & 0.30$^{+0.22}_{-0.16}$ & FUV$-$NUV, NUV$-$g, u$-$g, g$-$r, r$-$i, i$-$z \\ 
 37 & n4017 tidal & 179.73745 & 27.44398 & 4$^{+80}_{-3}$ & 0.16$^{+0.30}_{-0.16}$ & FUV$-$NUV, NUV$-$g \\ 
 38 & n4017 tidal & 179.74238 & 27.4396 & 1$^{+12}_{-0}$ & 0.00$^{+0.16}_{-0.00}$ & FUV$-$NUV, NUV$-$g, g$-$r \\ 
 39 & n4017 tidal & 179.71796 & 27.437119 & 7$^{+60}_{-2}$ & 0.00$^{+0.16}_{-0.00}$ & FUV$-$NUV, NUV$-$g, g$-$r, r$-$i \\ 
 40 & n4017 tidal & 179.7304 & 27.43336 & 1$^{+2}_{-0}$ & 0.00$^{+0.06}_{-0.00}$ & FUV$-$NUV \\ 
 41 & n4017 tidal & 179.69331 & 27.437551 & 7$^{+27}_{-2}$ & 0.00$^{+0.08}_{-0.00}$ & FUV$-$NUV, NUV$-$g, u$-$g, g$-$r, r$-$i \\ 
 42 & n4017 tidal & 179.66234 & 27.463605 & 7$^{+30}_{-1}$ & 0.00$^{+0.06}_{-0.00}$ & FUV$-$NUV, NUV$-$g \\ 
 43 & n4017 tidal & 179.7358 & 27.439398 & 43$^{+133}_{-38}$ & 0.22$^{+0.22}_{-0.22}$ & FUV$-$NUV, NUV$-$g, g$-$r, r$-$i \\ 
 44 & n4017 tidal & 179.71139 & 27.437124 & 8$^{+28}_{-7}$ & 0.30$^{+0.20}_{-0.16}$ & FUV$-$NUV, NUV$-$g, g$-$r \\ 
 45 & n4017 tidal & 179.70247 & 27.439421 & 9$^{+121}_{-2}$ & 0.38$^{+0.32}_{-0.32}$ & FUV$-$NUV, r$-$i \\ 
 
\hline \\ 
\multicolumn{7}{l}{\footnotesize{$^a$ We did not include upper limits in our fits}}\\ 
\end{tabular} 
\end{table}


\begin{thebibliography}{}

\bibitem[Abazajian et al.(2003)]{aba03}Abazajian, K, et al. 2003, AJ, 126, 2081

\bibitem[Arp(1966)]{arp66}Arp, H. 1966, Atlas of Peculiar Galaxies (Pasadena: Caltech)

\bibitem[Bastian(2005)]{bas05}Bastian, N. 2005, A\&A, 431, 905

\bibitem[Begum et al.(2008)]{beg08}Begum, A., Chengalur, J. N., Karachentsev, I.D., \& Sharina, M. E. 2008, MNRAS, 386, 138

\bibitem[Boissier et al.(2008)]{boi08}Boissier, S. et al. 2008, \apj, 681, 244

\bibitem[Bournaud, Duc, \& Emsellem(2008)]{bou08}Bournaud, F., Duc, P.-A., \& Emsellem, E. 2008, MNRAS, 289, 8

\bibitem[Bournaud \& Duc(2006)]{bou06}Bournaud, F. \& Duc, P. -A. 2006, A\&A, 456, 481

\bibitem[Boyce et al.(2001)]{boy01}Boyce, P. J. et al. 2001, \apj, 560, 127

\bibitem[Bushouse(1987)]{bus87}Bushouse, H. 1987, ApJ, 320, 49

\bibitem[Calzetti, Kinney, \& Storchi-Bergmann(1994)]{cal94}Calzetti, D., Kinney, A. L., \& Storchi-Bergmann, T. 1994, \apj, 429, 582

\bibitem[Cervi\~{n}o \& Valls-Gabaud(2003)]{cer03}Cervi\~{n}o, M. \& Valls-Gabaud, D. 2003, MNRAS, 338, 481

\bibitem[Dahari(1985)]{dah85}Dahari, O. 1985, ApJS, 57, 643 

\bibitem[de Mello et al.(2008)]{dem08}de Mello, D. F., Smith, L. J., Sabbi, E., Gallagher, J. S., Mountain, M., \& Harbeck, D. R. 2008, \aj, 135, 548

\bibitem[Duc, Bournaud, \& Boquien(2006)]{duc06}Duc, P. -A., Bournaud, F., \& Boquien, M. 2006, in IAU Symp. 237, ed. B. G. Elmegreen, \& J. Palous, 323 

\bibitem[Elmegreen et al.(2006)]{elm06}Elmegreen, D. M., Elmegreen, B. G., Kaufman, M., Sheth, K., Struck, C., Thomasson, M., \& Brinks, E. 2006, ApJ, 642, 158 
\bibitem[Elmegreen, Kaufman, \& Thomasson(1993)]{elm93}Elmegreen, B. G., Kaufman, M., \& Thomasson, M. 1993, ApJ, 412, 90

\bibitem[Fall, Chandar, \& Whitmore(2005)]{fal05}Fall, S.M., Chandar, R., \& Whitmore, B. C. 2005, ApJ, 631, L133

\bibitem[Gil de Paz et al.(2005)]{gil05}Gil de Paz A., et al., 2005, ApJ, 627, L29

\bibitem[Giroux et al.(2009)]{gir09}Giroux, M. L. et al. 2009, in preparation

\bibitem[Hancock et al.(2008)]{han08}Hancock, M., Smith, B. J., Giroux, M. L., \& Struck, C. 2008, MNRAS, 389, 1470

\bibitem[Hancock et al.(2007)]{han07}Hancock, M., Smith, B. J., Struck, C., Giroux, M. L., Appleton, P. N., Charmandaris, V., \& Reach, W. T. 2007, \aj, 133, 676

\bibitem[Higdon, Higdon \& Marshall(2006)]{hig06}Higdon, S. J. U., Higdon, J. L., \& Marshall, J. 2006, \apj, 640, 783 

\bibitem[Holtzman et al.(1992)]{hol92}Holtzman, J. A., et al. 1992, AJ, 103, 691

\bibitem[Holtzman et al.(1996)]{hol96}Holtzman, J. A., et al. 1996, AJ, 112, 416

\bibitem[Hunter \& Elmegreen(2004)]{hun04}Hunter, D. A. \& Elmegreen, B. G. 2004, AJ, 128, 2170

\bibitem[Jester et al.(2005)]{jes05}Jester, S. et al. 2005, \aj, 130, 873

\bibitem[Kaufman et al.(1997)]{kau97}Kaufman, M., Brinks, E., Elmegreen, D. M., Thomasson, M., Elmegreen, B. G., Struck, C., \& Klaric, M. 1997, \aj, 114, 2323 

\bibitem[Kroupa(2002)]{kro02}Kroupa, P. 2002, Science, 295, 85 

\bibitem[Leitherer et al.(1999)]{lei99}Leitherer, C., et al. 1999, \apjs, 123, 3

\bibitem[Lynden-Bell \& Jin(2008)]{lyn08}Lynden-Bell, D., \& Jin, S. 2008, MNRAS, 386, 245

\bibitem[Maybhate et al.(2007)]{man07}Maybhate, A., Masiero, J., Hibbard, J. E., Charlton, J. C., Palma, C., Knierman, K. A., \& English, J. 2007, MNRAS, 381, 59

\bibitem[Martin et al.(2005)]{mar05}Martin, D. C., et al. 2005, ApJ, 619, L1

\bibitem[Muller et al.(2004)]{mul04}Muller, E., Stanimirovi\'{c}, S., Rosolowsky, E., \& Staveley-Smith, L. 2004, \apj, 616, 845 

\bibitem[Muller et al.(2003)]{mul03}Muller, E., Staveley-Smith, L., Zealey, W., \& Stanimirovi\'{c}, S. 2003, MNRAS, 339, 105 

\bibitem[Neff et al.(2005)]{nef05}Neff, S. G., et al. 2005, ApJ, 619, L91

\bibitem[O'Connell, Gallagher, \& Hunter(1994)]{oco94}O'Connell, R. W., Gallagher, J. S., \& Hunter, D.A. 1994, ApJ, 433, 65 


\bibitem[Oke(1990)]{oke90}Oke, J., B., 1990, \aj, 99, 1621

\bibitem[Pisano et al.(2005)]{pis05}Pisano, D. J., Koo, D. C., Willmer, C. N. A., Noeske, K. G., \& Phillips, A. C. 2005, \apj, 630, L25

\bibitem[Press et al.(1992)]{pre92}Press, W. H., Teukolsky, S. A., Vetterling, W. T., \& Flannery, B. P. 2002, Numerical Recipes in Fortran:  the art of scientific computing, 2nd ed. (Cambridge University Press) 

\bibitem[Pryor \& Meylan(1993)]{pry93}Pryor, C., \& Meylan, G. 1993, in ASP conf. Series, Vol. 50, Structure and Dynamics of Globular Clusters, ed. S. G. Djorgovski \& G. Meylan (San Fransisco: ASP), 357

\bibitem[Recchi et al.(2006)]{rec06}Recchi, S., Theis, C., Kroupa, P., \& Hensler, G. 2007, A\&A, 470, 5

\bibitem[Romero-G\'{o}mez et al.(2008)]{rom08}Romero-G\'{o}mez, M., Athanassoula, J. J., \& Garc\'{i}a-G\'{o}mez, C. 2008, astro-ph/0801.3412

\bibitem[Schweizer et al.(1996)]{sch96}Schweizer, F., Miller, B. W., Whitmore, B. C., \& Fall, S. M. 1996, AJ, 112, 1839

\bibitem[Smith et al.(2008)]{smi08}Smith, B. J., et al. 2008, \aj, 135, 2406

\bibitem[Smith \& Struck(2001)]{smi01}Smith, B. J. \& Struck, C. 2001, AJ, 121, 710

\bibitem[Struck(1997)]{str97}Struck, C. 1997, ApJS, 113, 269

\bibitem[Struck(1999)]{str99}Struck, C. 1999, Phys Rep, 321, 1

\bibitem[Struck(2006)]{str06}Struck, C. 2006, AJ, 131, 1347


\bibitem[Struck \& Smith(2003)]{str03}Struck, C., \& Smith, B. J. 2003, ApJ, 589, 157

\bibitem[Sun et al.(2005)]{sun05}Sun, W. -H. et al. 2005, \apj, 630, 133

\bibitem[Tarchi et al.(2005)]{tar05}Tarchi, A., Ott, J., Pasquali, A., Ferraras, I., Castangia, P., \& Larsen, S. S. 2005, A\&A, 133, 136

\bibitem[Thilker et al.(2005)]{thi05}Thilker D.A., et al., 2005, ApJ, 619, L79

\bibitem[van Moorsel(1983)]{moo83}van Moorsel, G. A. 1983, A\&AS, 54, 19

\bibitem[V\'{a}zquez \& Leitherer(2005)]{vaz05}V\'{a}zquez, G. A. \& Leitherer, C. 2005, ApJ, 621, 695

\bibitem[Watson et al.(1996)]{wat96}Watson, A. M., et al. 1996, AJ, 112, 534

\bibitem[Whitmore \& Schweizer(1995)]{whi95}Whitmore, B. C. \& Schweizer, F. 1995, AJ, 109, 960 

\bibitem[Whitmore et al.(1993)]{whi93}Whitmore, B. C., Schweizer, F., Leitherer, C., Borne, K., \& Robert, C. 1993, AJ, 106, 1354 


\end{thebibliography}
\end{document}